\documentclass[aps,pre,showpacs,floats,twocolumn,floats,superscriptaddress,floatfix,groupedaddress]{revtex4-1}
\usepackage{bm}
\usepackage{url}
\usepackage{times}
\usepackage{verbatim}
\usepackage{theorem}
\usepackage{makeidx}
\usepackage{amsmath}
\usepackage{siunitx}
\usepackage{xy}
\usepackage{amssymb}
\usepackage{graphicx}
\usepackage{color}
\usepackage{epsfig}
\usepackage{subfigure}  % use for side-by-side figures
\usepackage[breaklinks]{hyperref}

\begin{document}
\title{Statistics of Lagrangian Trajectories in a Rotating Turbulent Flow}
\author{Priyanka Maity}
\email{priyanka.maity@icts.res.in}
\affiliation{International Centre for Theoretical Sciences, Tata Institute of Fundamental Research, Hessaraghatta, Hobli, Bangalore - 560089, India} 
\author{Rama Govindarajan}
\email{rama@icts.res.in}
\affiliation{International Centre for Theoretical Sciences, Tata Institute of Fundamental Research, Hessaraghatta, Hobli, Bangalore - 560089, India} 
\author{Samriddhi Sankar Ray}
\email{samriddhisankarray@gmail.com}
\affiliation{International Centre for Theoretical Sciences, Tata Institute of Fundamental Research, Hessaraghatta, Hobli, Bangalore - 560089, India} 
\begin{abstract}

We investigate the Lagrangian statistics of three-dimensional rotating turbulent flows 
through direct numerical simulations. We find that the emergence of coherent vortical 
structures because of the Coriolis force leads to a suppression of the ``flight-crash'' 
events reported by Xu, {\it et al.} [Proc. Natl. Acad. Sci. (U.S.A) {\bf 111}, 7558 (2014)]. 
We perform systematic studies to trace the origins of this suppression in the emergent geometry of 
the flow and show why such a Lagrangian measure of irreversibility may fail in the presence 
of rotation. 

\end{abstract}

\maketitle

The irreversibility of fully developed, homogeneous and isotropic turbulence,
as well as the non-trivial spatio-temporal structure of its (Eulerian) velocity
field shows up in an interesting way in the statistics of the kinetic energy
along Lagrangian trajectories. Xu, \textit{et al.}~\cite{Xu_2014}, measured the
kinetic energy of a tracer along its trajectory, as a function of time, to
show that the gain in kinetic energy (over time) is gradual whereas the loss is
rapid. (The average energy, statistically, is of course constant over time.)
This behaviour of the energy fluctuations is quantified most conveniently by
the statistics of energy increments (gain or loss) at small, but fixed, time
intervals. In the limiting case, the rate of change of 
the kinetic energy, or power, serves as a useful probe to understand how Eulerian irreversibility
manifests itself in the Lagrangian framework. Bhatnagar, \textit{et
al.}~\cite{bhatnagar_2018}, extended this idea to the case of heavy, inertial
particles, preferentially sampling the flow, to disentangle the effects of
irreversibility and flow geometry.

This feature of Lagrangian trajectories, dubbed as ``flight-crash events''~\cite{Xu_2014}, 
is a consequence of the dissipative nature of turbulent flows as well as the spatial 
structure of the Eulerian field with its intense, though sparse, regions of vorticity and 
more abundant, though milder, regions of strain. However, so far, measurements have 
been confined only to flows which are statistically homogeneous and isotropic. 
Therefore it is natural to ask if \textit{flight-crash} events are just as ubiquitous 
in turbulent settings with anisotropy and structures different from those seen 
in statistically homogeneous, isotropic turbulence. An obvious candidate for this 
is fully developed turbulent flows under rotation~\cite{Greenspan_1968,
Moffat_1983, davidson_2013} which are seen in a variety of processes spanning 
scales ranging from the astrophysical~\cite{Barnes_2001,James_2008,Reun_2017},
geophysical~\cite{Aurnou_2015} to the industrial~\cite{Dumitrescu_2004}.  In
all these phenomena, although the Coriolis force does no work, it leads to the
formation of large-scale columnar vortices leading to dynamics quite different
from non-rotating, three-dimensional flow. In particular, 
rotation gives rise to an enhanced accumulation of energy in modes perpendicular to
the plane of rotation~\cite{Smith_1999, Muller_2007, Minnini_2009}, an inverse
energy cascade in 3D turbulence~\cite{Bartello_1995,metais_1996, Yarom_2013},
generation of inertial waves~\cite{Bewley_2007,davidson_2013, Yarom_2017}, and
an increase in length scales parallel to the axis of
rotation~\cite{ibbetson_tritton_1975}. Consequently, rotating turbulence has 
been the subject of much  experimental~\cite{ibbetson_tritton_1975,
hopfinger_1982, Bewley_2007, Morize_2005, moisy_2011, Yarom_2013, Yarom_2017} and 
theoretical~\cite{bartello_1994,Bartello_1995,metais_1996,
godeferd_1999,Zeman_1994, hattori_2004, Muller_2007, Minnini_2009,
sreenivasan_2008, Pouquet_2010} investigations in the last few decades.

A striking effect of rotation is on the geometry of the flow. Rapid rotation
leads to a two-dimensionalisation of the flow through the formation of columnar
(cyclonic) vortices parallel to the rotation axis~\cite{Proudman_1916,
Taylor_1917, Greenspan_1968, davidson_2013, Luca_2016} as well as an emergent
anisotropy through the breaking of the cyclone-anticylone symmetry.  This
effect, characterised and measured in experiments~\cite{hopfinger_1982,
Morize_2005, moisy_2011, Gallet_2014} and direct numerical simulations
(DNS)~\cite{bartello_1994, godeferd_1999, sreenivasan_2008}, stems from an
enhanced (cyclonic) vortex stretching because of the Coriolis force. 

The effect of these emergent two-dimensional vortical structures in a three-dimensional  
flow on Lagrangian measurements has received attention only recently~\cite{Castello_2011, 
Castello_PRL_2011, Castello_PRE_2011, Luca_2016}. In particular, Biferale, \textit{et al.}~\cite{Luca_2016}, through 
state-of-the-art DNSs explored these consequences on the mixing and transport 
properties of particles (both tracers and inertial) in rotating turbulence. However, the 
effect of such coherent structures on individual Lagrangian (tracer) trajectories from 
the point of view of time-irreversibility remains an open one.

In this paper, we investigate this aspect of rotating turbulence and find that
time-irreversibility, in the Lagrangian sense~\cite{Xu_2014,bhatnagar_2018},
decreases as the effect of rotation, and hence columnar vortices, becomes
stronger. These results are rationalised by careful measurements of the
correlation between the topology of the flow and the tracer trajectories which
suggests that the spatial structure of a flow is critical in determining the
strength of \textit{flight-crashes} even if the flow itself retains the same
degree of irreversibility through a finite dissipation. 

We begin with the three-dimensional Navier-Stokes equation 
\begin{equation}
\frac{\partial {\bf u}}{\partial t}+({\bf u}\cdot\nabla){\bf u} + 2 ({\bf \Omega} \times {\bf u}) = -\nabla P' +\nu \nabla^2{\bf u}+ {\bf f}
\label{ns}
\end{equation}
for the velocity field ${\bf u}$ of a unit-density fluid rotating about a fixed
axis with a rate $\Omega$, along with the
incompressibility condition $\nabla\cdot {\bf u} = 0$. We use an external
forcing ${\bf f}$, on wavenumber(s) $k_f$, to drive the fluid (with kinematic viscosity $\nu$) to a
statistically steady state associated with an energy (viscous) dissipation rate $\epsilon$. 
The pressure  $P' =  P_0 - \frac{1}{2} \vert {\bf \Omega} \times {\bf r} \vert ^2$ 
absorbs the centrifugal contribution from the rotating frame along with the natural pressure $P_0$ in the fluid in 
absence of rotation.

\begin{table*}
{
\setlength{\tabcolsep}{3pt}
\begin{tabular}{|ccccccccccccccc|}
%\hline
%\hline
%$N$  & $\nu$ & $\delta t$ & $N_p$ & $\lambda$ &$Re_{\lambda}$ & $\eta$ & $\tau_\eta$&$\epsilon$ & $\tau_{\eta}/\delta t$ & $\eta/dx$ &$k_{\rm max} \eta$ &   $\Omega$ & $Ro$ & $\alpha$ \\
%\hline
%$512$  & $10^{-3}$ & $4 \times 10^{-4}$ & $10^6$ & $0.12$& $90$ & $6 \times 10^{-3}$& $3.33 \times 10^{-2}$ &$0.89$ & $86$ & $0.6$ & $2.56$ & $0 - 2.0$ & $\infty - 0.06$ &$5 \times 10^{-3}$ \\
%\hline
%\hline
%
\hline
\hline
$N$  & $\delta t$ & $\nu$ & $\alpha$ & $N_p$ & $\epsilon$ & $\eta$ & $\tau_\eta$ & $\lambda$ & $Re_{\lambda}$ &
$\tau_{\eta}/\delta t$ & $\eta/dx$ & $k_{\rm max} \eta$ & $\Omega$ & $Ro$ \\
\hline
$512$  & $4 \times 10^{-4}$ & $10^{-3}$ & $5 \times 10^{-3}$ &$10^6$ & $0.89$ & $6 \times 10^{-3}$ & $3.33 \times 10^{-2}$ &
$0.12$ & $90$ & $86$ & $0.6$ & $2.56$ & $0 - 2.0$ & $\infty - 0.06$ \\
\hline
\hline

\end{tabular}

\caption{Parameters for the simulations in a $2\pi$ periodic cube: $N$ is number of collocation
	points in each direction,  $\delta t$ is the time step of integration, $\nu$ is the kinematic viscosity of the
	fluid, $\alpha$ is the coefficient of the large scale friction, and $N_p$ represent the
	number of tracer particles seeded in the flow. The mean energy dissipation rate $\epsilon = 2 \nu \sum\limits_k k^2 E(k)$, while 
	$\eta = \left( \frac{\nu^3}{\epsilon} 	\right)^{1/4}$ and $\tau_{\eta} = \left( \frac{\nu}{\epsilon} 	\right)^{1/2}$ are the Kolmogorov length and time scales, respectively. The Taylor length scale of the flow is denoted by  $\lambda = \sqrt{\frac{15 \nu_{\rm rms}^2}{\epsilon}}$ 
	(where, $u_{\rm rms}$ is the root-mean-squared fluid velocity) and $Re_{\lambda} = u_{rms} \lambda/ \nu$ represents the Reynolds number
	 corresponding to the Taylor microscale. The number of collocation points determine the grid spacing $dx = 2\pi/N$ and  the maximum wave number $k_{\rm max}$ of the simulations. The rotation rate $\Omega$ defines the Rossby number $Ro = \frac{u_{rms}}{2 L \Omega}$. We choose
	  seven different strengths of rotation 	rates $\Omega = 0, 0.1, 0.5, 0.75, 1.0$, and $2.0$ yielding Rossby numbers 
	$Ro = \infty, 1.23, 0.24, 0.16, 0.12, 0.08$, and $0.06$, respectively. }
\label{tab:parameters}
}
\end{table*}

Apart from the Reynolds number, rotational turbulent flows in a box of size $L$
(in our case, $L=2\pi$), with typical room-mean-square velocities $u_{\rm
rms}$, are conveniently characterized by a second dimensionless number, the
Rossby number $Ro \equiv {u_{\rm rms}/(2 L \Omega)}$, which is the ratio of the
inertial to the Coriolis term. Furthermore, the additional Coriolis term leads
to a natural scale-separation, the so-called Zeman wavenumber $k_{\Omega} \sim
\sqrt{\frac{\Omega^3}{\epsilon}}$, which sets the scale where the local fluid
turnover time ($\epsilon^{-1/3}k^{-2/3}$) is of the same order as
$\Omega^{-1}$. For strongly rotating flows ($Ro \ll 1$) and wavenumbers $k <
k_\Omega$, the kinetic energy spectrum $E(k) = |u_k|^2$ tends to steepen
leading to a scaling $E(k) \sim
k^{-2}$~\cite{Zhou_1995,Baroud_1998,Yeung_1998,hattori_2004, Luca_2016} while
retaining the usual Kolmogorov spectrum $E(k) \sim k^{-5/3}$ at higher
wavenumbers. 

We solve the Navier-Stokes equation in the rotating frame (Eq.~\ref{ns})
through the standard pseudo-spectral method~\cite{canuto_book}, with a
second-order Adams-Bashforth scheme for time-marching, in a $2\pi$-periodic
periodic cubic box with the axis of rotation being the $z$-axis. We use $N^3 =
512^3$ collocation points and an external constant energy injection force,
acting on wavenumbers $k \leq 3$, to drive the system to a statistically
steady state with a Taylor-scale Reynolds number $Re_\lambda \approx 100$.  
As is common in such numerical simulations, we introduce a small
additional frictional term in the form of an inverse Laplacian, with a small
coefficient $\alpha = 0.005$ to damp out the energy
which piles up at the smaller modes due to the inverse cascade set in motion by
the rotation. { The details of the simulation are presented in Table~\ref{tab:parameters}.
}

\begin{figure}
\includegraphics[width=1.0\columnwidth]{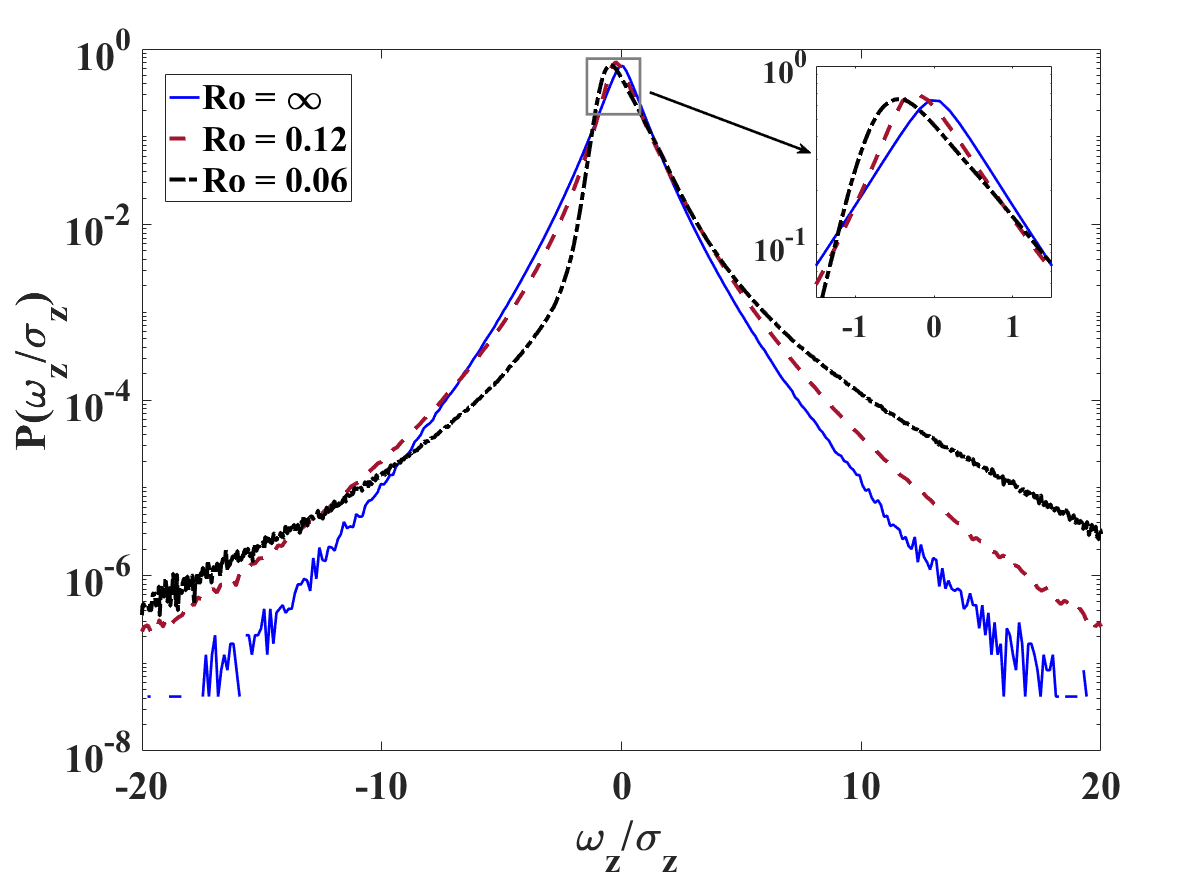}
\caption{Probability distribution functions of the vertical component of the vorticity $\omega_z$, normalized by its standard deviation 
$\sigma_z$ for $Ro = \infty$ (blue solid curve), $0.12$ (red dashed curve), and $0.06$ (black dashed-dotted curve)}
\label{fig:omega_pdf}
\end{figure} 

We begin by addressing the question of how the reorganization of the flow, in the presence of a Coriolis force, 
influences the trajectories of tracers, which sample the phase space of the flow uniformly. 
{We therefore seed, randomly and homogeneously, our flow, once it has reached a statistically stationary state, with 
$N_p = 10^6$ tracer particles with initial velocities identical to the velocity of the fluid at particle position.
The dynamics of a single tracer, with a trajectory ${\bf r}_p$, is given by:}
 \begin{eqnarray}
	 \frac{d{\bf r}_p}{dt} = {\bf v}_p; \quad {\bf v}_p = {\bf u}({\bf r}_p).  
	 \label{tracer_vel}
\end{eqnarray}
Numerically, we use a trilinear interpolation scheme to obtain the fluid velocity ${\bf u}({\bf r}_p)$ at particles positions, since these are typically off-grid. 

\begin{figure}
 \includegraphics[width=1.0\columnwidth]{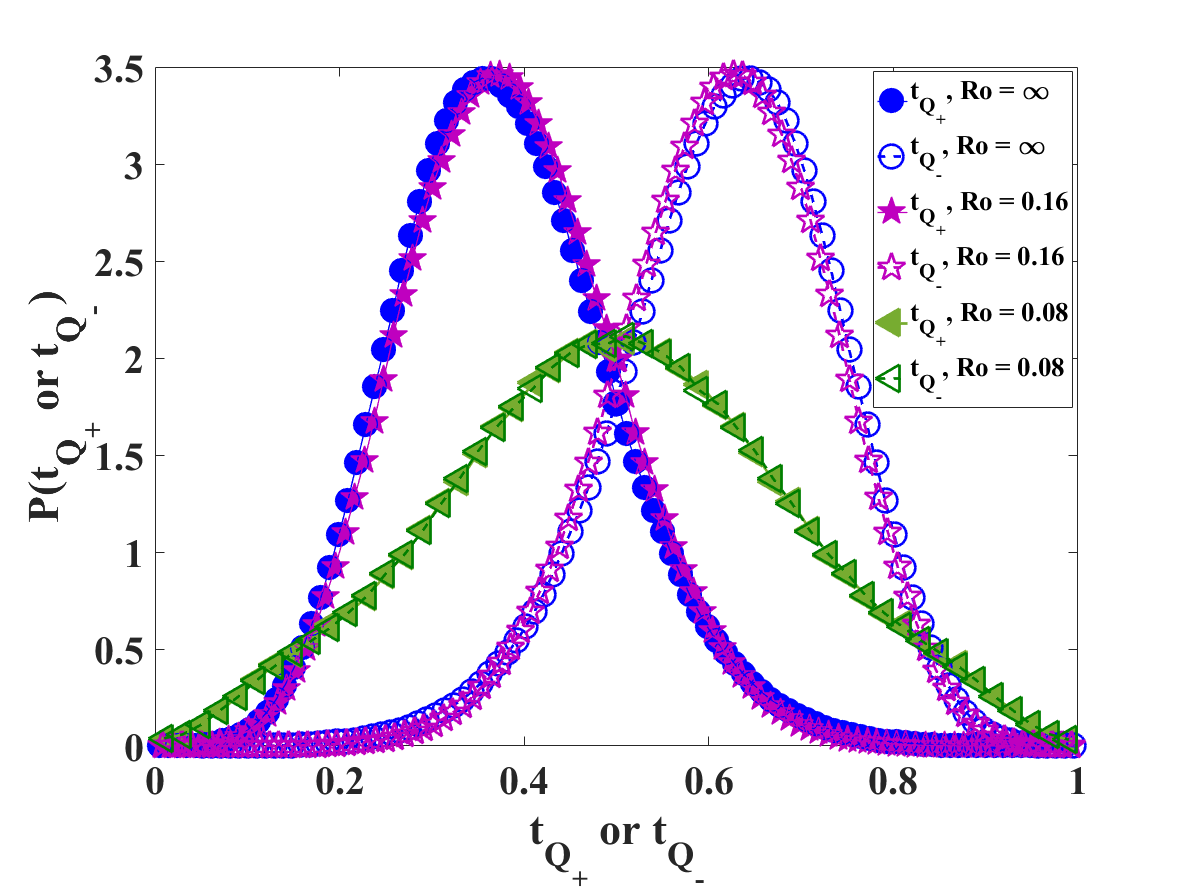}
 \caption{Probability distribution functions of the fraction of time particles spend in 
	vortical ($t_{{\rm Q}_+}$) (filled markers) and straining ($t_{{\rm Q}_{-}}$) (open markers) regions for 
$Ro = \infty$ (circles), $0.16$ (stars), and $0.08$ (triangles).}
\label{fig:q_time_dist}
\end{figure}

For such steady states in the presence of rotation, as the Rossby 
number decreases, the emergence of strong coherent cyclonic vortices---due to enhanced stretching of cyclonic vortices and destabilization of anti-cyclonic vortices---leads to an increased positive skewness in the probability distribution function 
(pdf) of the vorticity component $\omega_z$ in the direction $z$ of the axis of rotation. In Fig.~\ref{fig:omega_pdf} we show this distribution function, measured along the Lagrangian trajectories of the particles, 
for different values of $Ro$. Our observation, that the pdf becomes increasing skewed, is 
consistent with that seen in experiments of rotating flows, such as 
by Morize, {\it et al.}~\cite{Morize_2005}. Furthermore, these distributions show an exponential tail 
and peak (Fig.~\ref{fig:omega_pdf}, inset) at mildly negative values of $\omega_z$ (with decreasing $Ro$) ensuring an overall positive skewness 
for $Ro \ll 1$. 

A convenient way to measure the correlation between the structure---broadly vortical and straining---of the flow 
and the Lagrangian dynamics is through the 
second invariant of the velocity gradient tensor $\nabla {\bf u}$~\cite{Hunt_1988,jeong_hussain_1995}:
\begin{equation}
{\rm Q} = \frac{1}{2}({\parallel  {\bf \Theta} \parallel}^2 - {\parallel {\bf \Sigma} \parallel}^2),
\label{eq:Q}
\end{equation}
{
where ${\parallel {\bf \Theta} \parallel} = {\rm Tr}\big[{\bf \Theta} {\bf
\Theta}^T]^{1/2}$ and ${\parallel {\bf \Sigma} \parallel} = {\rm Tr}\big[{\bf \Sigma}
{\bf \Sigma}^T]^{1/2}$; the superscript $T$ denotes the transpose of a matrix and
Tr its trace. The symmetric component of the velocity gradient tensor, or the rate-of-strain tensor, 
is given by ${\bf \Sigma} = \frac{1}{2} \left[\nabla {\bf u} + (\nabla {\bf u})^T \right]$ and 
the  anti-symmetric component by
${\bf \Theta} = \frac{1}{2} \left[\nabla {\bf u} - (\nabla {\bf u})^T \right]$. 
Hence, $\rm Q$ gives a local measure of the relative strengths of these two tensors.
(Note that the non-standard notation $\Theta$ and $\Sigma$ are
used because the symbols $\Omega$  and $S$ are taken up for the mean rotation rate of the system 
and the symmetry function, respectively.)} Such a local measurement of $\rm Q$ is thus a useful diagnostic to 
determine if the flow at any point, in an Eulerian framework, is dominated by vortices (${\rm Q} \geq 0$) 
or by straining regions (${\rm Q} < 0$). Similarly this `$\rm Q$' criterion can be applied in Lagrangian measurements, 
such as ours, by measuring the $\rm Q$ value of the flow seen by a Lagrangian particle along its trajectory.

With this formalism, we begin by investigating how the emergence of coherent
columnar vortices, with decreasing Rossby numbers, leads to a bias in the
Lagrangian sampling of the flow, and hence to the flight-crash picture. We
begin by calculating the fraction of time spent by the tracers in vortical
$t_{{\rm Q}_{+}}$ and straining $t_{{\rm Q}_{-}}$ regions. {This
is done most conveniently by measuring $\rm Q$ along each tracer trajectory which
allows us to calculate, for each trajectory, the fraction of time it has spent
in vortical $t_{{\rm Q}_{+}}$ (${\rm Q} \geq 0$) or straining $t_{{\rm Q}_{-}}$
(${\rm Q} < 0$) regions; from the data of these times for $N_p$, we are able to
construct the pdf of these residence times.} In Fig.~\ref{fig:q_time_dist} we
show the distribution of these times for different Rossby numbers (including
the case of no rotation).  We find that for weak ($Ro=0.16$) or no
($Ro=\infty$) rotation, Lagrangian particles spend a disproportionately large
fraction of time in strain-dominated regions as compared to vorticity dominated
ones (as seen by the blue and magenta curves in Fig~\ref{fig:q_time_dist}).
{This is because for no (or weak) rotation, the flow is
characterized by  weaker but spatially extended straining regions in contrast
to the more localised and sparse regions of strong vorticity. Hence the tracers
spend more time in straining regions than in vortical ones.  As $Ro \to 0$, the flow
re-organizes itself with a proliferation of extended vortical structures.
Consequently, the fraction of the flow with positive ${\rm Q}$ becomes
comparable to that with negative ${\rm Q}$ and tracers spend more
time in vortical regions than they would be if the effect of rotation is weak.  In fact, for $Ro = 0.08$ it can be seen (green curves
in Fig~\ref{fig:q_time_dist}) that the distribution of $t_{{\rm Q}_+}$ and
$t_{{\rm Q}_-}$ are now practically identical.}  It is important to recall
that Bhatnagar, {\it et al.},~\cite{Bhatnagar_2016} showed an apparently
contrary behaviour for their Lagrangian measurements in a non-rotating flow.
This is due to their use of the $\rm \Delta$-criterion~\cite{chong_1990} which
over-samples the vortices by considering regions which have {\it small}
negative values of $\rm Q$. We have checked that our results are consistent with
Bhatnagar, {\it et al.},~\cite{Bhatnagar_2016} when we use the
$\rm \Delta$-criterion and not the $\rm Q$-criterion. 

\begin{figure}
\includegraphics[width=1.0\columnwidth]{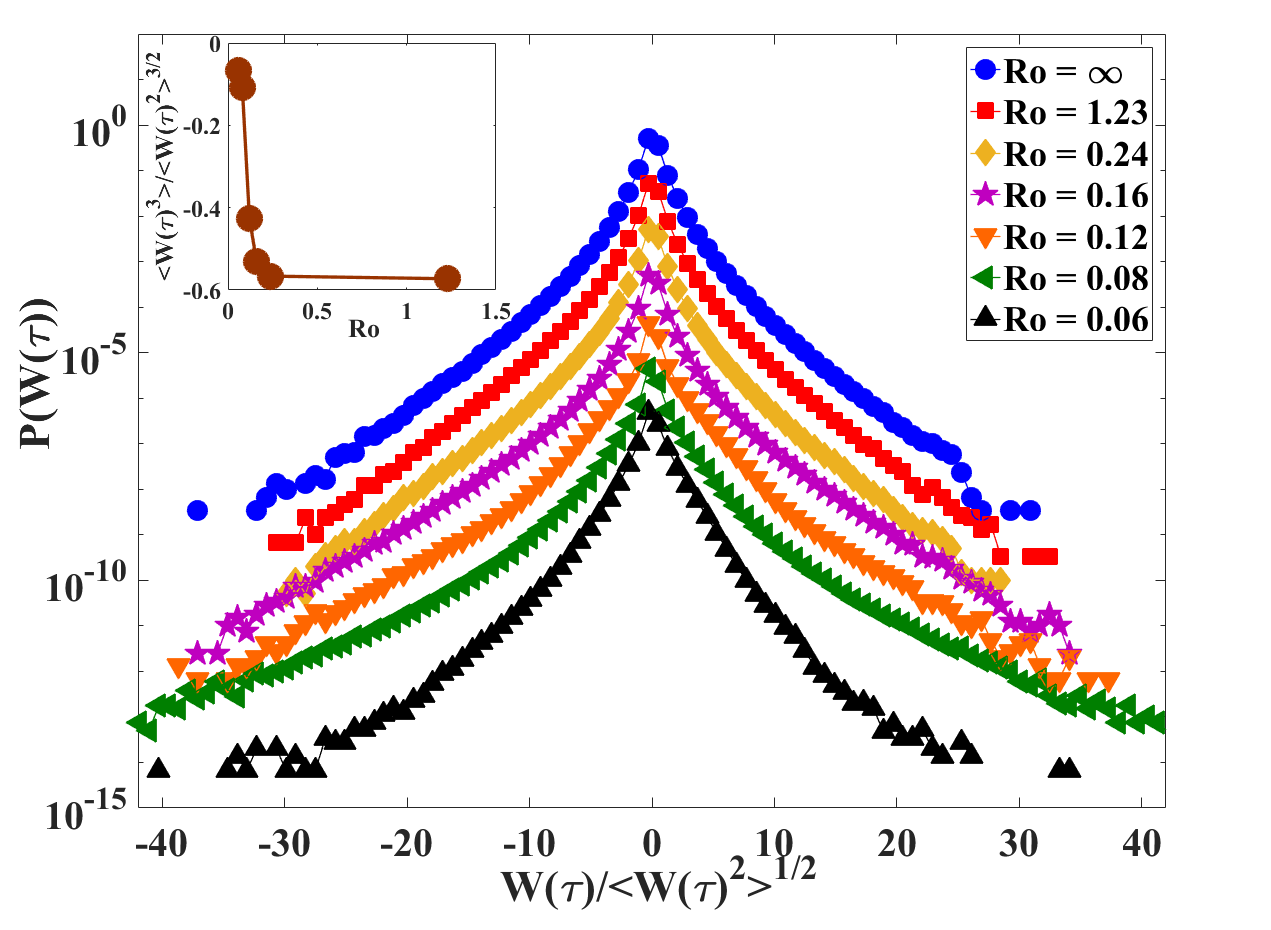}
\caption{Probability distribution function of the energy increment $W(\tau) = E(t + \tau) - E(t)$, normalised by its root-mean-square value 
$<W(\tau)^2>^{1/2}$ for $\tau/\tau_{\eta}=2$. The curves are for different values of $Ro$ (see legend) and artificially shifted 
by factors of 10 for clarity. { (Inset) The skewness of the pdfs of $W(\tau)$ as a function of $Ro$}.}
\label{fig:Wpdf} 
\end{figure}
This striking feature of the residence times of tracers in different regions of
the flow, as a function of the Rossby number, leads us to ask if it plays a
role in negating flight-crashes as a useful probe for irreversibility.  We
begin by measuring the probability distribution function of the Lagrangian
energy increments $W(\tau) = E(t+\tau) - E(t)$, where $E(t)$ is the kinetic
energy of the tracer at any time $t$. In non-rotating
flows~\cite{Xu_2014,bhatnagar_2018}, this distribution is negatively skewed
because gains $W(\tau)>0$ in energy are slower than their dips $W(\tau)<0$ for
any fixed $\tau$. { In Fig.~\ref{fig:Wpdf}, we plot this distribution, along with 
the skewness (see inset) as a function of $Ro$, at time
$\tau/\tau_\eta = 2.0$, for several different values of the Rossby number
(artificially separated for clarity). We see, and quantify through the inset in Fig.~\ref{fig:Wpdf}, 
a less skewed behaviour of the energy increments as $Ro \to 0$. (It should be noted that the choice of 
$\tau/\tau_\eta = 2.0$  is arbitrary; the pdfs of the energy increments are always negatively skewed 
but this is more pronounced when the time-increments are small.)}
 
\begin{figure}
\includegraphics[width=1.0\columnwidth]{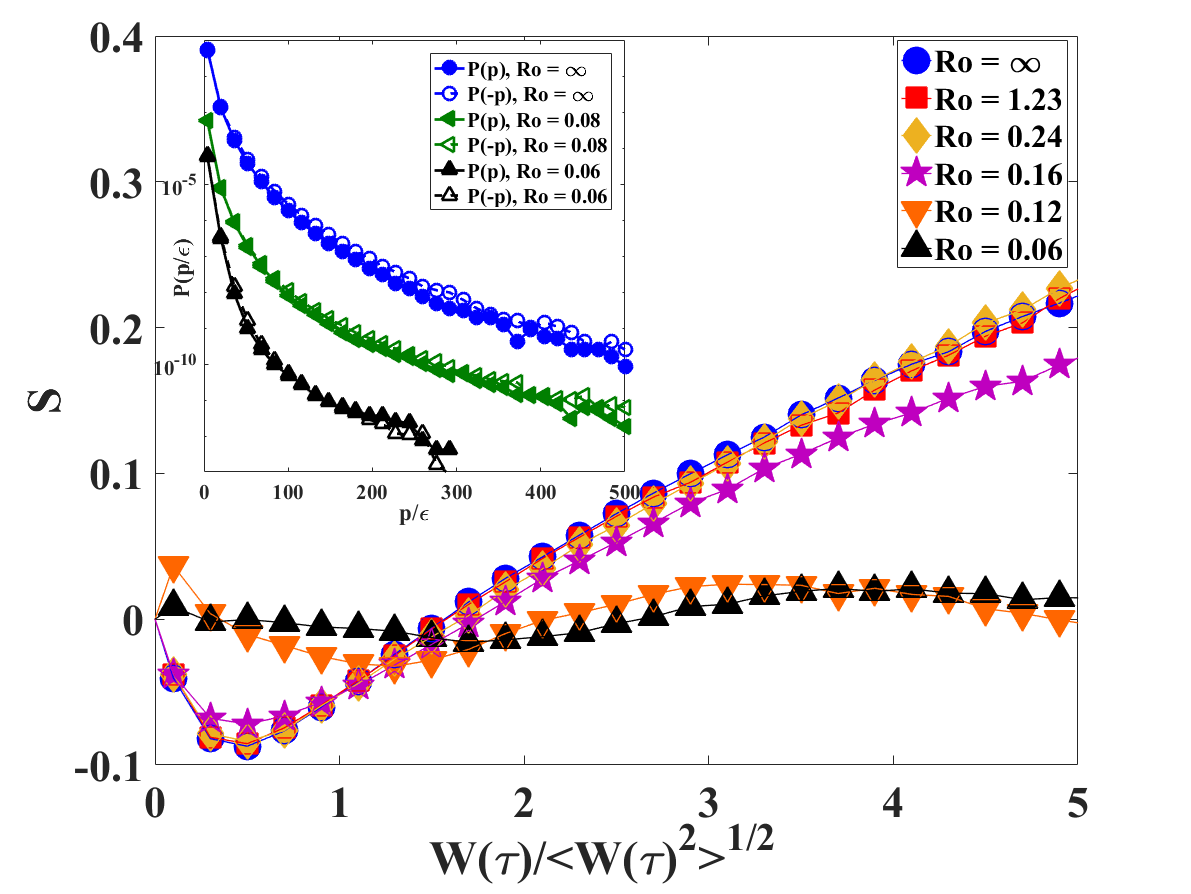}
\caption{Representative plots of the symmetry function S versus the energy increment $W(\tau)$ normalised by 
 its root-mean-square value $<W(\tau)^2>^{1/2}$ for $\tau/\tau_{\eta}=2$ for different values of $Ro$ (see legend).
Inset: Probability distribution functions of the negative (open symbols) and positive (filled symbols) values of the 
Lagrangian power for  $Ro = \infty$, $0.08$, and $0.06$; the negative tail has been reflected for ease of comparison.}
\label{fig:S} 
\end{figure}

\begin{figure*}
\includegraphics[width=1.0\linewidth]{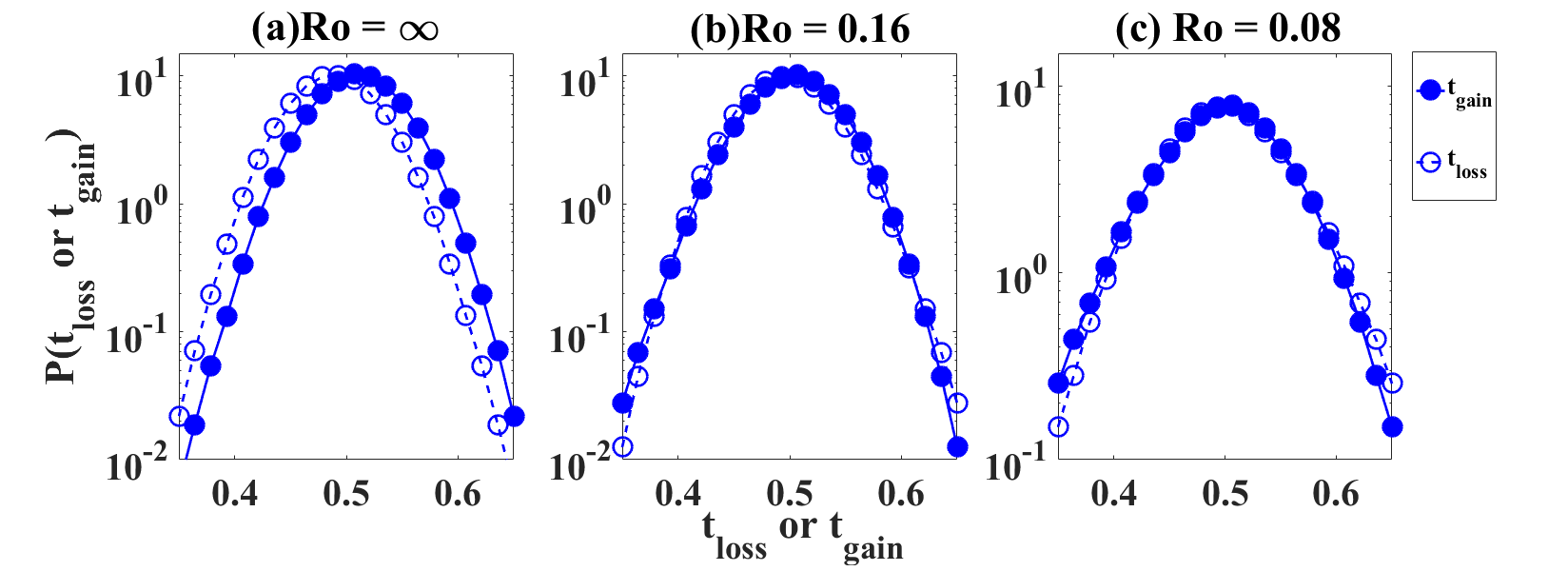}
\caption{Probability distribution functions of the fraction of time during which tracers gain $t_{\rm gain}$ (filled circles) and
lose $t_{\rm loss}$ (open circles) energy for (a) $ Ro = \infty$, (b) $ Ro = 0.16$, and (c) $ Ro = 0.08$.}
\label{fig:p_time_dist}
\end{figure*}
A natural interpretation of such distributions are made in the light of 
fluctuation-dissipation theorems which, at least for simpler systems 
which can be modelled as being coupled to thermostats, states that~\cite{Ciliberto_2010} 
\begin{equation} 
S \equiv \ln \left [\frac{P(-W)}{P(W)}\right] \propto W,
\label{eq:FT} 
\end{equation}
where $S$, as seen, is a symmetry function constructed from the ratio of energy jump probabilities.
Xu, \textit{et al.}~\cite{Xu_2014}, found their measurements to be strongly 
fluctuating and hence found no convincing evidence that 
such a function $S$ actually scales linearly with $W$. However, in our simulations, 
shown in Fig.~\ref{fig:S}), given the volume of data and statistics, 
a plot of the symmetry function $S$ versus the normalized $W$ shows a much cleaner trend.
We see a small window of nearly-linear behavior at moderate values of $W(\tau)$ for cases of no or negligible rotation. On the other hand, when the Rossby number 
is very small, the plot of $S$ is essentially flat and lies close to $0$. This, then, is the first clear evidence 
that the skewness in the distribution of $W$ diminishes as the flow reorganizes itself 
under strong rotation. Thus a combination of emergent coherent vortices and inverse cascade 
in strongly rotating flows---while remaining dissipative---seems to negate the possibility of flight-crash 
events along Lagrangian trajectories as a probe for measuring the irreversibility of the flow. 
This conclusion is further strengthened in the inset of Fig.~\ref{fig:S} where we plot the (suitably normalized) 
distributions of the positive and negative values of the power $p \equiv {dE/dt}$ for different 
values of the Rossby number. For no or negligible rotations, it is visible that at higher power, the curve for negative power lies above that for positive power (reflective of the irreversibility and flight-crashes~\cite{Xu_2014}); however when $Ro \ll 1$, the two pdfs are 
hardly distinguishable.

\begin{figure}
\includegraphics[width=1.0\columnwidth]{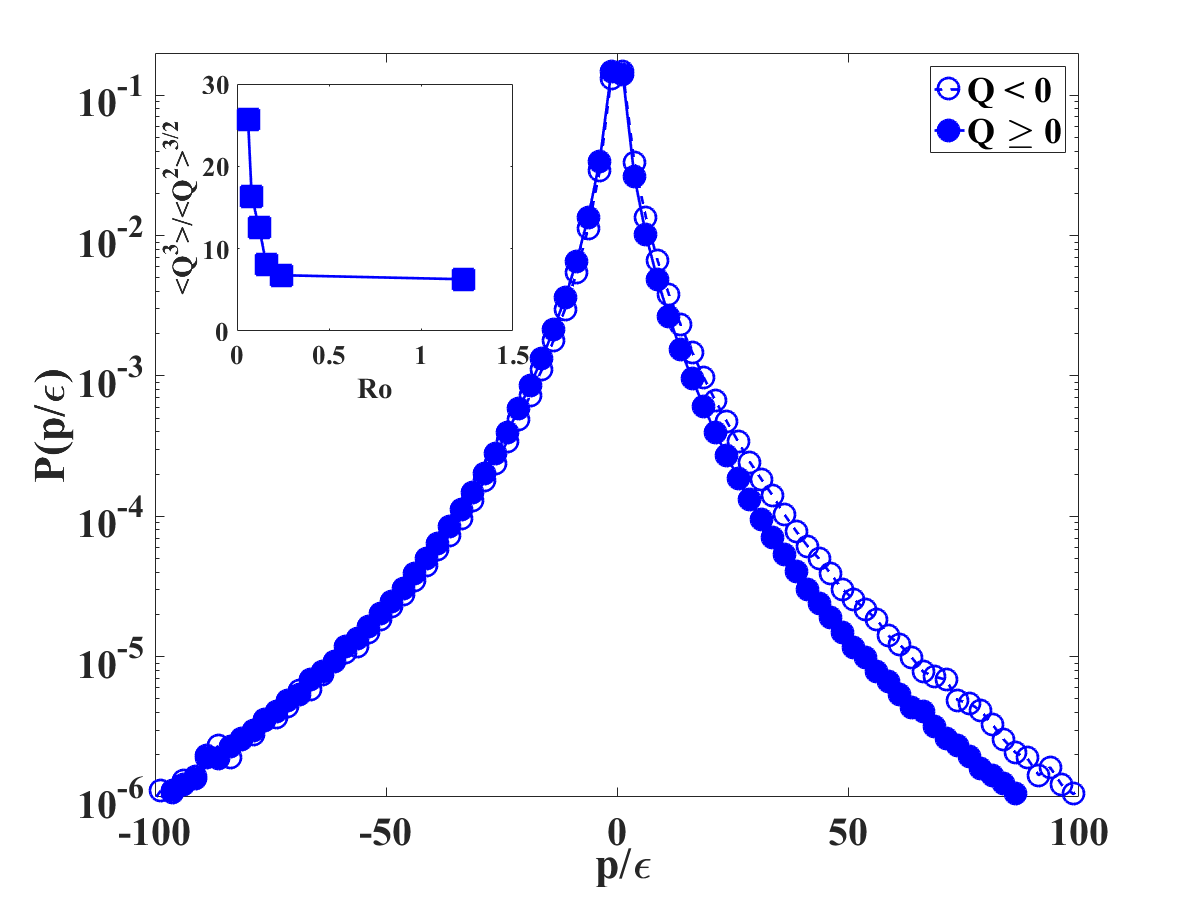}
\caption{Representative plots, for $Ro = 0.16$, of the pdf of the Lagrangian power, normalized by energy dissipation rate $\epsilon$, 
conditioned on whether the particles are in vortical ${\rm Q} \geq 0$ (filled circles) or straining regions ${\rm Q} < 0$ (open circles). (Inset) 
The skewness of the ${\rm Q}$ showing a sharp increase with decreasing Rossby number.}
\label{fig:power_pdf_Q}
\end{figure}
Since the macroscopic dissipation is held constant in our calculations, this
lack of flight-crashes must stem from the emergent anisotropic geometry of the
flow under rotation. To quantify this, we measure the power $p$ as a function
of time for our $N_p$ Lagrangian trajectories from which it is possible to
construct the fraction of time that a particle spends in gaining ($p>0$) or
losing ($p<0$) energy. For homogeneous, isotropic turbulence, the proliferation
for flight-crash events would suggest that the fraction of time $t_{\rm gain}$
spent in gaining energy must be larger than the fraction of time $t_{\rm loss}$
spent in losing it. In Fig.~\ref{fig:p_time_dist}(a) we plot the distribution
of both $t_{\rm gain}$ and $t_{\rm loss}$ to find evidence for this: The
(Gaussian) distribution of the time for energy gain is shifted to the right
compared to the one for energy loss. However, as rotation starts to
dominate, the distribution for both start becoming identical with a mean
fraction of time spent for either gaining or losing energy being half
(Figs.~\ref{fig:p_time_dist} (b) and (c)). { This variation of the pdfs of 
$t_{\rm gain}$ and $t_{\rm loss}$ follow a trend similar to, but not exactly,  as the distributions of the residence times
in Fig.~\ref{fig:q_time_dist}. This is because 
the collapse of pdfs of $t_{{\rm Q}_{+}}$ and $t_{{\rm Q}_{-}}$ onto each other, as $Ro \to 0$,
is much more dramatic when compared to the gradual merging of the pdfs of $t_{\rm gain}$ and $t_{\rm loss}$ for similar changes in $Ro$.
This is expected, of course, because unlike a direct measurement of the residence times, the 
fraction of times spent in losing or gaining kinetic energy, although correlated to 
where it resides in the flow, is not uniquely determined by the local flow geometry.}

We also measure the distribution of $p$ along Lagrangian trajectories
\textit{and} conditioned on whether they are in vortical $({\rm Q} \ge 0)$ or in
straining $({\rm Q} < 0)$ regions. In Fig.~\ref{fig:power_pdf_Q} we plot the pdf of
the power, conditioned on the geometry of the flow for $Ro = 0.16$, and see
evidence that in straining regions, energy gains are more probable than energy losses, whereas in vortical
regions the probabilities are similar. Now, as rotation in increased, the fraction of the flow in vortical regions is higher. Moreover these vortices are stronger and more coherent on an average. This can be seen in the Lagrangian skewness in the distribution of ${\rm Q}$ (Fig.~\ref{fig:power_pdf_Q}, inset). Therefore at higher rotation rates, gains in energy are as frequent, and have the same distribution, as losses.

\begin{figure}
 \includegraphics[width=1.0\columnwidth]{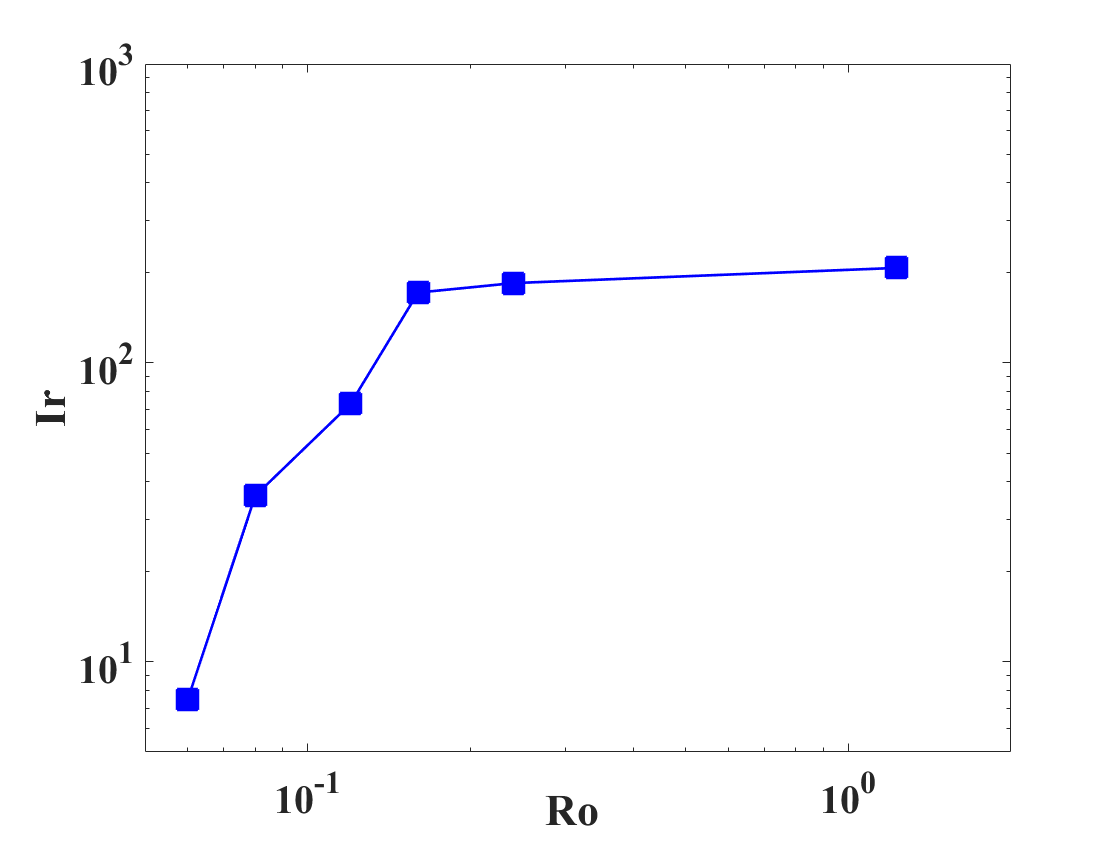}
\caption{The irreversibility Ir as a function of $Ro$ showing a sharp increase, by an order of magnitude, as $Ro \to 1$.}
\label{fig:Ir} 
\end{figure}

What does all of this mean for the central question of this work, namely Lagrangian irreversibility and its 
connections with the geometry of the flow? { In statistically steady state, Lagrangian reversibility or 
time reversibility should imply that the probability of energy gain ($p > 0$) or energy loss ($p < 0$)  should be equal. Therefore, any asymmetry in the pdfs of $p$ would suggest the breaking of Lagrangian or time 
reversibility in the system.} One way to measure this irreverisibility, is to consider the 
quantity~\cite{Xu_2014}:

 \begin{equation}
 {\rm Ir} = \frac{-<p^3>}{\epsilon^3}.  
  \label{eq:irreversibility}
  \end{equation}
For homogeneous and isotropic turbulence, $Ir \gg 1$ and increases with the Reynolds number of the flow. This 
stems from the fact that ``flight-crashes'' proliferate in such flows with increasing Reynolds numbers leading 
to an ever-increasing skewness in the distribution of the power $p$. We have, however, seen that because of the 
Coriolis force, the flow reorganizes leading to, e.g., a depletion in the skewness of $\rm Q$ along Lagrangian trajectories. 
Could the effect be as strong in measurements of $Ir$? In Fig.~\ref{fig:Ir} we plot
Ir as a function of $Ro$ and find a sharp decrease as soon as $Ro < 1$. Indeed
for $Ro \ll 1$, the decrease in the irreversibility is  by an order of
magnitude compared to the case where the rotation is negligible. (We have
checked that our value of Ir for $\Omega = 0$ is consistent with the findings
in Ref.~\cite{Xu_2014}.) 

Indeed, in recent times, this issue of reversibility has been re-examined in a
variety of problems which range from the use of the time-reversible
Navier-Stokes equation~\cite{Shukla2018} to how the suppression of small-scale
intermittency through Fourier-decimation~\cite{PRL-dec,Ray-review} lead to an
emergent reversibility when measure via Lagrangian Lyapunov
exponents~\cite{Ray-Irr}.  Our study is however different from these. Unlike
the use of a fluctuating thermostat to replace the usual viscosity in the
Navier-Stokes equation leading to time-reversibility or the suppression of a
subset of triadic interactions to solve the equations on a quenched, disordered
lattice, we show that even for the \textit{true} equations of motion, rotation
and the consequent emergence of coherent, anisotropic structures is enough to
alter the statistics of Lagrangian trajectories.  In particular, at high
rotation rates, the notion of flights and crashes cease to exist and thus does
not allow an easy interpretation of time-irreversibility in terms of such
Lagrangian probes. We hope that this work, which does not rely on modifications 
to the equations of hydrodynamics, will lead to experiments designed to 
look at this specific aspect of Lagrangian irreversibility in experiments in future. 

We thank J. R. Picardo for useful suggestions and discussions { and H. J. H. Clercx for several 
useful references}. A part of this
work was facilitated by a program organized at ICTS: \textit{Turbulence from
Angstroms to Light Years} (ICTS/Prog-taly2018/01).  The simulations were
performed on the ICTS clusters {\it Mowgli}, {\it Tetris} and {\it Mario} as
well as the work stations from the project ECR/2015/000361: {\it Goopy} and
{\it Bagha}.  SSR acknowledges DST (India) project ECR/2015/000361 for
financial support. 

\bibliographystyle{apsrev4-1}.
\bibliography{ref}
\end{document}